\documentclass[aps,floatfix]{revtex4}
\usepackage{graphicx}
\begin{document}
\title{Colloquium on the Higgs Boson}
\author{Philip D. Mannheim}
\affiliation{Department of Physics, University of Connecticut, Storrs, CT 06269, USA.
email: philip.mannheim@uconn.edu}
\date{February 24, 2016}
\begin{abstract}
In 2013 the Nobel Prize in Physics was awarded to Francois Englert and
Peter Higgs for their development in 1964 of the mass generation mechanism
(the Higgs mechanism) in local gauge theories. This mechanism requires the
existence of a massive scalar particle, the Higgs boson, and in 2012 the
Higgs boson was finally observed at the Large Hadron Collider after being sought for almost half a century. In this pedagogical article  we review the work of these
Nobel recipients and discuss its implications. We approach the topic from
the perspective of a dynamically generated  Higgs boson that is a fermion-antifermion
bound state rather than an  elementary field that appears in an input
Lagrangian. In particular, we emphasize the connection with the  BCS theory of superconductivity. We identify the double-well Higgs potential not as a fundamental potential but as  a mean-field effective Lagrangian
with a dynamical Higgs boson being generated through  a
residual interaction that accompanies the mean-field Lagrangian. We discuss what we believe to be
the key challenge raised by the discovery of the Higgs boson, namely determining
whether it is elementary or composite, and suggest that the width of the Higgs boson might serve as a suitable diagnostic for discriminating between an elementary Higgs boson and a composite one.
\end{abstract}
 
\maketitle

\tableofcontents

\bigskip

\section{Introduction}
\subsection{Preamble}

The 2013 Nobel Prize in Physics was awarded to Francois Englert and Peter Higgs for their work in 1964 that led in 2012 to the discovery  at the CERN Large Hadron Collider (LHC) of the Higgs Boson after its being sought for almost 50 years. It is great tragedy that Robert Brout, the joint author with Francois Englert of one of the papers that led to the 2013 Nobel Prize, died in 2011, just one year before the discovery of the Higgs boson and two years before the awarding of the Nobel prize for it. (For me personally this is keenly felt since my first post-doc was with Robert and Francois in Brussels 1970 - 1972.) 

The paper coauthored by Englert and Brout appeared on August 31, 1964 in Physical Review Letters \cite{Englert1964} after being submitted on June 26, 1964 and was two and one half pages long. Higgs wrote two papers on the topic. His first paper appeared on September 15, 1964 in Physics Letters \cite{Higgs1964a}  after being submitted on July 27, 1964 and was one and one half pages long, and the second paper appeared on October 19, 1964 in Physical Review Letters \cite{Higgs1964b} after being submitted on August 31, 1964 and was also one and one half pages long. Thus a grand total of just five and one half pages.
 
The significance of the Higgs mechanism introduced in these papers and the Higgs Boson identified by Higgs is that they are tied in with the theory of the origin of mass, and of the way that mass can arise through collective effects (known as broken symmetry) that only occur in systems with a large number of degrees of freedom. Such collective effects are properties that a system of many objects collectively possess that each one individually does not -- the whole being greater than the sum of its parts. A typical example is temperature. A single molecule of $H_2O$ does not have a temperature, and one cannot tell if it was taken from ice, water or steam. These different phases are collective properties of large numbers of $H_2O$ molecules acting in unison. Moreover, as one changes the temperature all the $H_2O$ molecules can act collectively to change the phase (freezing water into ice for instance), with it being the existence of such phase changes that is central to broken symmetry.

I counted at least 19 times that Nobel Prizes in Physics have in one way or another been given for aspects of the problem: 
Dirac (1933); Anderson (1936); Lamb (1955); Landau (1962); Tomonaga, Schwinger, Feynman (1965); Gell-Mann (1969); Bardeen, Cooper, Schrieffer (1972); Richter, Ting (1976); Glashow, Salam, Weinberg (1979); Wilson (1982); Rubbia, van der Meer (1984); Friedman, Kendall, Taylor (1990); Lee, Osheroff, Richardson (1996); 't Hooft,Veltman (1999); Abrikosov, Ginzburg, Leggett  (2003); Gross, Politzer, Wilczek (2004); Nambu, Kobayashi, Maskawa (2008); Englert, Higgs (2013); Kajita, McDonald (2015).
And this leaves out Anderson who made major contributions to collective aspects of mass generation and Yang who (with Mills) developed non-Abelian Yang-Mills gauge theories but got Nobel prizes (Anderson 1977, Yang 1957) for something else. 

In this paper  I provide a pedagogical review the work of the work that led up to the discovery of the Higgs boson 
and discuss its implications \cite{footnote1}.

\subsection{Ideas About Mass}

As introduced by Newton mass was mechanical. The first ideas on dynamical mass were due to Poincare (Poincare stresses needed to stabilize an electron all of whose mass  came from its own electromagnetic self-energy according to $mc^2=e^2/r$). However this was all classical.

With quantum field theory, the mass of a particle is able to change through self interactions (radiative corrections to the self-energy -- Lamb shift) to give $m=m_0+\delta m$, or through a change in vacuum (Bardeen, Cooper, Schrieffer -- BCS theory)  according to $E=p^2/2m+\Delta$ where $\Delta$ is the self-consistent gap parameter. 
Then through Nambu (1960) and Goldstone (1961)  the possibility arose that not just some but in fact all of the mass could come from self interaction, and especially so for gauge bosons, viz. Anderson (1958, 1963), Englert and Brout  (1964), Higgs (1964), Guralnik, Hagen, and Kibble (1964). This culminated in the Weinberg (1967), Salam (1968), and Glashow (1961, 1970)  renormalizable $SU(2)\times U(1)$ local gauge theory of electroweak interactions, and the confirming discoveries first of weak neutral currents (1973), then charmed particles (1974), then the intermediate vector bosons of the weak interactions (1983), and finally the Higgs boson (2012). All of this is possible because of Dirac's Hilbert space formulation of quantum mechanics in which one sets $\psi(x)=\langle x|\psi\rangle$, with the physics being in the properties of the states $|\psi\rangle$. We thus live in Hilbert space and not in coordinate space, and not only that, there is altogether more in Hilbert space than one could imagine, such as half-integer spin and collective macroscopic quantum systems such as superconductors and superfluids. In this Hilbert space we find an infinite Dirac sea of negative energy particles. This large number of degrees of freedom can act collectively to provide the dynamics needed to produce mass generation and the  Higgs boson.

\section{The Higgs Boson Discovery}

The discovery of the Higgs boson was announced by CERN on July 4, 2012, accompanied by simultaneous announcements by the experimental groups ATLAS and CMS at the Large Hadron Collider at CERN, as then followed by parallel publications that were simultaneously submitted to Physics Letters B on July 31, 2012 and published on September 17, 2012, one publication by the ATLAS Collaboration \cite{ATLAS2012}, and the other by the CMS Collaboration \cite{CMS2012}. The ATLAS paper had 2924 authors, and the CMS paper had 2883. The Higgs boson signature used by both collaborations was to look for the decay into lepton pairs of any Higgs boson that might be produced in high energy  proton proton collisions at the Large Hadron Collider.

From amongst a set of 
$10^{15}$ proton-proton collisions produced at the Large Hadron Collider, of the order of 240,000 collisions produce a Higgs boson. Of them just 350 decay into pairs of gamma rays and just 8 decay into a pair of leptons. 
The search for the Higgs boson is thus a search for some very rare events. Thus to see Higgs bosons one needs an energy high enough to produce them and the sensitivity to see such rare decays when they are produced. In searches over the years it was not known in what energy regime to look for Higgs particles, with the Large Hadron Collider proving to be the collider whose energy was high enough that one could finally explore in detail the 125 GeV energy domain where the Higgs boson was ultimately found to exist.

\section{Background Leading to the Higgs Mechanism and Higgs Boson Papers in 1964}

In order to characterize macroscopic ordered phases in a general way Landau introduced the concept of a macroscopic order parameter $\phi$. For a ferromagnet for instance $\phi$ would represent the spontaneous magnetization $M$ and would thus be a matrix element of a field operator in an ordered quantum state that described the ordered magnetic phase. Building on this approach Ginzburg and Landau \cite{Ginzburg1950} wrote down a Lagrangian for such a $\phi$ for a superconductor, with kinetic energy $\vec{\nabla}\phi\cdot \vec{\nabla}\phi/2$ and potential $V(\phi)=\phi^4/4-(T_C-T) \phi^2/2$, where $T_C$ is the critical temperature.  For temperatures above the critical temperature the potential would have the shape of a single well, viz. like the letter U, with the coefficient of the $\phi^2$ term being positive. For temperatures below the critical temperature  the potential would have the shape of a double well, viz. like the letter W, with the coefficient of the $\phi^2$ term being negative. Above  the critical temperature the order parameter would be zero at the minimum  of the potential  (normal phase with state vector $|N\rangle$ in which $\langle N|\phi|N\rangle=0$). Below the critical temperature the order parameter would be nonzero (superconducting state $|S\rangle$ in which $\langle S|\phi|S\rangle=(T_C-T)^{1/2}$ is nonzero).

In 1957 Bardeen, Cooper, and Schrieffer  \cite{Bardeen1957} developed a macroscopic theory of superconductivity (BCS)  based on Cooper pairing of electrons  in the presence of a filled Fermi sea of electrons, and explicitly constructed the state $|S\rangle$. In this state the matrix element $\langle S|\psi(x)\psi(x)|S\rangle$ was equal to a spacetime independent  function $\Delta$, the gap parameter, which led to a mass shift to electrons propagating in the superconductor of the form $E=p^2/2m+\Delta$. The gap parameter $\Delta$ would be temperature dependent ($\sim (T_C-T)^{1/2}$), and would only be nonzero below the critical temperature. In 1959 Gorkov \cite{Gorkov1959} was able to derive the Ginzburg-Landau Lagrangian starting from the BCS theory  and identify the order parameter as $\phi(x)=\langle C|\psi(x)\psi(x)|C\rangle$ where $|C\rangle$ is a coherent state in the Hilbert space based on $|S\rangle$. In the superconducting case then $\phi$ is not itself a quantum-field-theoretic operator (viz. a q-number  operator that would have a canonical conjugate with which it would not commute) but is instead a c-number matrix element  of a q-number field operator in a macroscopic coherent quantum state.

In 1958 Anderson \cite{Anderson1958} used the BCS theory to explain the Meissner effect, an effect in which electromagnetism becomes short range inside a superconductor, with photons propagating in it becoming massive. The effect was one of spontaneous breakdown of local gauge invariance, and was explored in detail by Anderson \cite{Anderson1958} and Nambu \cite{Nambu1960a}.

In parallel with these studies Nambu \cite{Nambu1960b},  Goldstone \cite{Goldstone1961}, and Nambu and Jona-Lasinio \cite{Nambu1961} explored the spontaneous  breakdown of some continuous global symmetries and showed that collective massless excitations (Goldstone bosons) were generated, and that the analog gap parameter would provide for dynamically induced fermion masses. In 1962  Goldstone, Salam, and Weinberg \cite{Goldstone1962} showed that there would always be massless Goldstone bosons in any Lorentz invariant theory in which a continuous global symmetry was spontaneously broken. While one could avoid this outcome if the symmetry was also broken in the Lagrangian, as was thought to be the case for the pion, a non-massless but near Goldstone particle (i.e. one with broken symmetry suppressed couplings to matter at low energies), in general the possible presence of massless Goldstone bosons was a quite problematic outcome because it would imply the existence of non-observed long range forces.  

In 1962 Schwinger \cite{Schwinger1962a,Schwinger1962b} raised the question of whether gauge invariance actually required that photons be massless, and noted for the inverse photon propagator $D^{-1}(q^2)=q^2-q^2\Pi(q^2)$  that if the vacuum polarization  $\Pi(q^2)$ had a massless pole of the form $\Pi(q^2)=m^2/q^2$, then $D^{-1}(q^2)$  would behave as the massive particle $D^{-1}(q^2)=q^2-m^2$. A massless Goldstone boson could thus produce a massive vector boson. 

With Anderson having shown that a photon would become become massive in a superconductor, there was a spirited discussion in the literature (Anderson \cite{Anderson1963}, Klein and Lee \cite{Klein1964}, Gilbert \cite{Gilbert1964}) as to whether an effect such as this might hold in a relativistic  theory as well or whether it might just have been an artifact of the fact that the BCS theory was non-relativistic. With the work of Englert and Brout \cite{Englert1964} and Higgs \cite{Higgs1964a,Higgs1964b}, and then Guralnik, Hagen, and Kibble \cite{Guralnik1964}, the issue  was finally resolved, with it being established that in the relativistic case the Goldstone theorem did not in fact hold if there was a spontaneous breakdown of a continuous local theory, with the would-be Goldstone boson no longer being an observable massless particle but instead combining with the initially massless vector boson to produce a massive vector boson. Technically, this mechanism should be known as the Anderson, Englert, Brout, Higgs, Guralnik, Hagen, Kibble mechanism, and while it has undergone many name variations over time, it is now commonly called the Higgs mechanism. What set Higgs' work apart from the others was that in his 1964 Physical Review Letters paper Higgs noted that as well as a massive gauge boson there should also be an observable massive scalar boson, this being the Higgs boson.

At the time of its development  in 1964 there was not much interest in the Higgs mechanism, with all of the Englert and Brout, Higgs, and Guralnik, Hagen and Kibble papers getting hardly any citations during the 1960s at all. The primary reason for this was that at the time there was little interest in non-Abelian Yang-Mills gauge theories in general, broken or unbroken, and not only was there no experimental indication at all that one should consider them, it was not clear if Yang-Mills theories were even quantum-mechanically viable. All this changed in the early 1970s when 't Hooft and Veltman showed that these theories were renormalizable,  and large amounts of data started to point in the direction of  the relevance of non-Abelian gauge theories to physics, leading to the $SU(3)\times SU(2)\times U(1)$ picture of strong, electromagnetic and weak interactions, which culminated in the discoveries of the $W^{+}$, $W^{-}$ and $Z_0$ intermediate vector bosons (1983) with masses that were generated by the Higgs mechanism, and then finally the Higgs boson itself (2012). What gave the Higgs boson the prominence that it ultimately came to have was the realization that in the electroweak $SU(2)\times U(1)$  theory the Higgs boson not only gives masses to the gauge bosons while maintaining renormalizability, but through its Yukawa couplings to the quarks and leptons of the theory it gives masses to the fermions as well. The Higgs boson is thus  responsible not just for the masses of the gauge bosons then but for the masses of all the other fundamental particles as well, causing it to be dubbed the  ``god particle" \cite{footnote2}.

\section{Broken Symmetry}
\label{S0}
\subsection{Global Discrete Symmetry -- Real scalar field -- Goldstone (1961)}

\begin{figure}[htpb]
\begin{center}
\includegraphics[width=5.0in,height=3.0in]{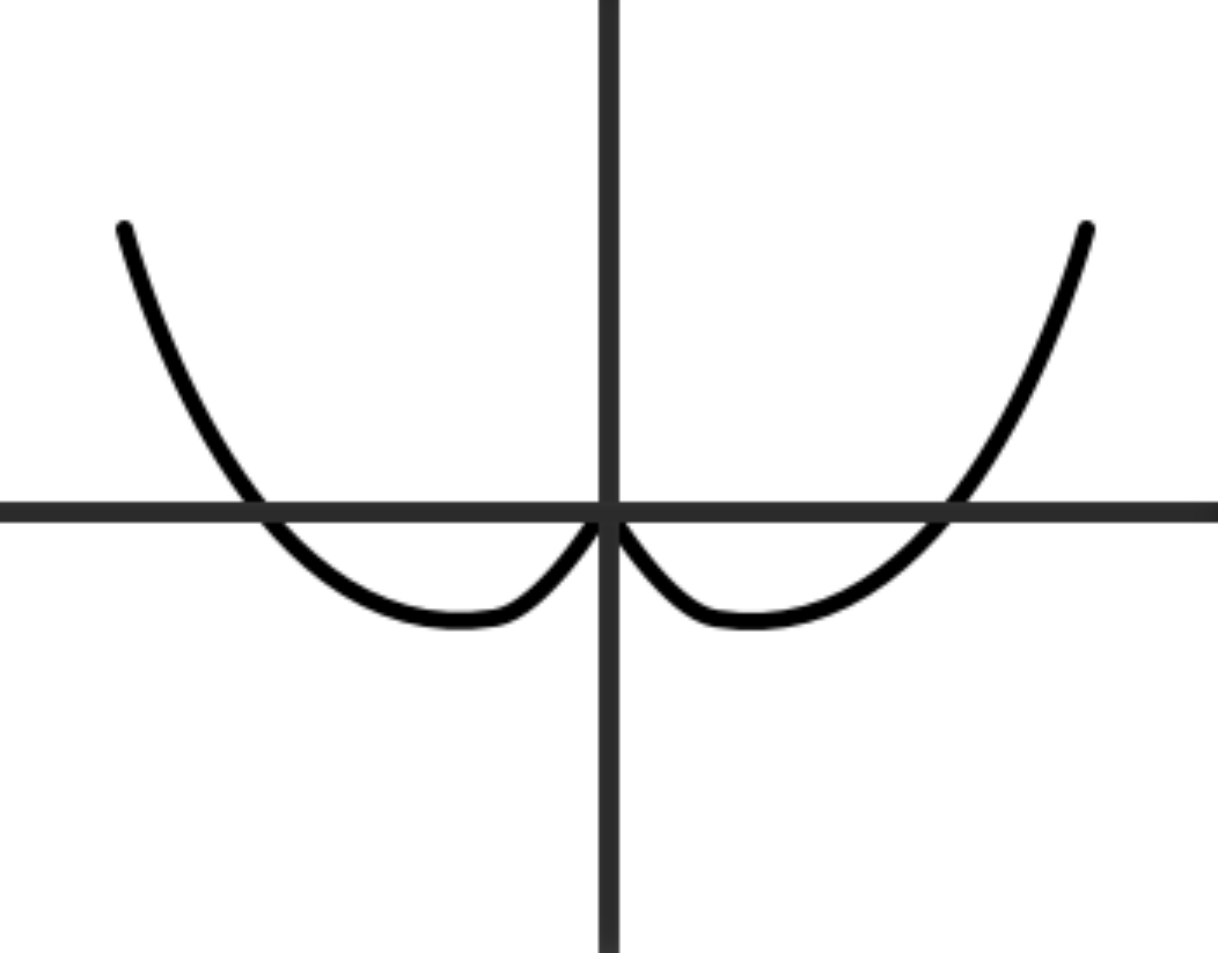}
\caption{Double Well Potential  -- $V(\phi)$ plotted as a function of  $\phi$}
\end{center}
\end{figure}
Even though broken symmetry is an intrinsically quantum-mechanical phenomenon, one can illustrate many of  the aspects needed to understand the Higgs boson by considering properties not of a q-number quantum field such as a  quantum scalar field $\hat{\phi}(x)$ but of its c-number vacuum matrix element $\phi(x)=\langle \Omega|\hat{\phi}(x)|\Omega\rangle$. Below we shall explore the quantum mechanics  of broken symmetry, but for pedagogical purposes it is convenient to first proceed as though one could treat the problem classically by studying properties of the c-number $\phi(x)$ without needing to delve into its origin or significance.

Thus consider a real classical scalar field (just one degree of freedom) with a potential energy in the form of the  double-well potential shown in Fig. (1), viz. a potential shaped like a letter W with two wells:
\begin{equation}
V(\phi)=\frac{1}{4}\lambda^2\phi^4-\frac{1}{2}\mu^2\phi^2.
\label{M1}
\end{equation}
This potential has a discrete symmetry under $\phi \rightarrow -\phi$, with its first two derivatives being given by
\begin{equation}
\frac{dV(\phi)}{d\phi}=\lambda^2\phi^3-\mu^2\phi,~~~\frac{d^2V(\phi)}{d\phi^2}=3\lambda^2\phi^2-\mu^2.
\label{M2}
\end{equation}
The potential has a local maximum at $\phi=0$ where $V(\phi=0)$ is zero and $d^2V(\phi=0)/d\phi^2$ is negative, and two-fold {\bf degenerate} global minima at $\phi=+\mu/\lambda$  and $\phi=-\mu/\lambda$ where $V(\phi=\pm \mu/\lambda)$ is equal to $-\mu^4/4\lambda^2$ and $d^2V(\phi=\pm \lambda/\mu)/d\phi^2$ is positive. Since $\phi=0$ is a local maximum, if we consider small oscillations  around $\phi=0$ of the form $\phi =0+\chi$ we generate a negative quadratic term $-(1/2)\mu^2\chi^2$ and thus negative mass squared, viz. $m^2=-\mu^2$, a so-called tachyon. The tachyon signals an instability of the configuration with $\phi=0$ (i.e. we roll away from the top of the hill).

However if we fluctuate around either global minimum (i.e. we oscillate in the vertical around the  bottom of either hill) by setting $\phi=\pm \mu/\lambda +\chi$ we get 
\begin{equation}
V(\phi)=-\frac{\mu^4}{4\lambda^2}+\mu^2\chi^2\pm \mu\lambda\chi^3+\frac{1}{4}\lambda^2\chi^4.
\label{M3}
\end{equation}
The field $\chi$ now has the positive $m^2=+2\mu^2$ ($=-2\times  m^2({\rm tachyon})$). Thus (Goldstone  1961) the would-be tachyonic particle becomes  a massive  particle, and with one scalar field we obtain one particle. This particle is the {\bf Higgs boson} in embryo.

The $-\mu^4/4\lambda^2$ potential term contributes to the {\bf cosmological constant}, and would be unacceptably large ($10^{60}$ times too large) if the boson has the 125 GeV mass that the Higgs boson has now been found to have.

All this arises because the minimum is two-fold degenerate, and picking either one breaks the symmetry spontaneously, since while one could just as equally be in either one or the other, one could not be in both. This is just like people at a dinner. Each one can take the cup to their left or their right, but once one person has done so, the rest have no choice. However a person at the opposite end of the table may not know what choice was made at the other end of the table and may make the opposite choice of cup, and thus persons in the middle could finish up with no cup. To ensure that this does not happen we need long range correlations -- hence massless Goldstone bosons.

\subsection{Global Continuous Symmetry -- Complex scalar field -- Goldstone (1961)}

Consider a complex scalar field (two degrees of freedom) $\phi=\phi_1+i\phi_2=re^{i\theta}$, $\phi^*\phi=\phi_1^2+\phi_2^2=r^2$, with a potential energy in the shape of a rotated letter W or a broad-brimmed, high-crowned Mexican Hat:   

\begin{equation}
V(\phi)=\frac{1}{4}\lambda^2(\phi^*\phi)^2-\frac{1}{2}\mu^2\phi^*\phi
=\frac{1}{4}\lambda^2(\phi_1^2+\phi_2^2)^2-\frac{1}{2}\mu^2(\phi_1^2+\phi_2^2).
\label{M4}
\end{equation}
The potential has a continuous global symmetry of the form  $\phi \rightarrow e^{i\alpha} \phi$ with constant $\alpha$, with derivatives
\begin{eqnarray}
\frac{dV(\phi)}{d\phi_1}=\lambda^2\phi_1^3+\lambda^2\phi_2^2\phi_1 -\mu^2\phi_1,~~~~
\frac{dV(\phi)}{d\phi_2}=\lambda^2\phi_2^3+\lambda^2\phi_1^2\phi_2 -\mu^2\phi_2.
\label{M5}
\end{eqnarray}
This potential has a local maximum at $\phi_1=0,\phi_2=0$ where $V(\phi=0)$ is zero, and infinitely {\bf degenerate} global minima at $\phi_1^2+\phi_2^2=\mu^2/\lambda^2$ (the entire 360 degree valley or trough between the brim and the crown of the Mexican hat). Again we would have  a tachyon if we expand around the local maximum, only this time we would get two. So fluctuate around any one of the global minima by setting $\phi_1=\mu/\lambda +\chi_1$, $\phi_2=\chi_2$, to get 
\begin{equation}
V(\phi)=-\frac{\mu^4}{4\lambda^2}+\mu^2\chi_1^2+\mu\lambda\chi_1^3+\frac{1}{4}\lambda^2\chi_1^4+\mu\lambda\chi_1\chi_2^2+\frac{\lambda^2}{2}\chi_1^2\chi_2^2.
\label{M6}
\end{equation}
The (embryonic) Higgs boson field is now $\chi_1$ with $m^2=+2\mu^2$. However, the field $\chi_2$ no has {\bf no mass at all} (it corresponds to horizontal oscillations along the valley floor), and is known as a {\bf Goldstone boson}  (Goldstone 1961). Thus from a complex scalar field we obtain two particles. Since the Goldstone boson is massless, it travels at the speed of light. It is thus intrinsically relativistic, and being massless can provide for long range correlations. Moreover, if such particles exist then they could generate fermion masses  entirely dynamically (Nambu 1960, Nambu and Jona-Lasinio 1961), with the pion actually serving this purpose.

The $-\mu^4/4\lambda^2$ potential term remains and the cosmological constant problem is just as severe as before.

Now if we have massless particles, we would get long range forces (just like photons), and yet nuclear and weak forces are short range. So what can we do about such Goldstone bosons. Two possibilities - they could get some mass because the symmetry is not exact (pion mass), or we could get rid of them altogether (the Higgs mechanism).

\subsection{Local Continuous Symmetry -- Complex Scalar Field and Gauge Field -- Higgs (1964)}

Consider the model studied by Higgs (1964) consisting of a complex scalar field (two degrees of freedom) coupled to a massless vector gauge boson (another two degrees of freedom) for a total of four degrees of freedom in all, viz. $\phi=re^{i\theta}$, $\phi^*\phi=r^2$, $F_{\mu\nu}=\partial_{\mu}A_{\nu}-\partial_{\nu}A_{\mu}$. The system has  kinetic energy $K$ and potential energy $V$:
\begin{eqnarray}
K&=&\frac{1}{2}(-i\partial_{\mu}+eA_{\mu})(re^{-i\theta})(i\partial^{\mu}+eA^{\mu})(re^{i\theta})-\frac{1}{4}F_{\mu\nu}F^{\mu\nu}
\nonumber\\
&=&\frac{1}{2}\partial_{\mu}r\partial^{\mu}r+\frac{1}{2}r^2(eA_{\mu}-\partial_{\mu}\theta)(eA^{\mu}-\partial^{\mu}\theta)
-\frac{1}{4}F_{\mu\nu}F^{\mu\nu},
\nonumber \\
V(\phi)&=&\frac{1}{4}\lambda^2(\phi^*\phi)^2-\frac{1}{2}\mu^2\phi^*\phi
=\frac{1}{4}\lambda^2r^4-\frac{1}{2}\mu^2r^2,
\label{M7}
\end{eqnarray}
and because of the gauge boson the system is now invariant under continuous {\bf local} gauge transformations of the form $\phi \rightarrow e^{i\alpha(x)} \phi$, $eA_{\mu}\rightarrow eA_{\mu}+\partial_{\mu}\alpha(x)$ with spacetime dependent  $\alpha(x)$. With derivatives
\begin{equation}
\frac{dV(\phi)}{dr}=\lambda^2r^3-\mu^2r,\qquad \frac{d^2V(\phi)}{dr^2}=3\lambda^2r^2-\mu^2,
\label{M8}
\end{equation}
the potential has a local maximum at $r=0$ where $V(r=0)$ is zero and {\bf degenerate} global minima at $r=\mu/\lambda$ (infinitely degenerate since independent of $\theta$). Again we would have two tachyons if we expand around the local maximum. So fluctuate around the global minimum by setting $r=\mu/\lambda +\chi_1$, $\theta_2=\chi_2$. On defining $B_{\mu}=A_{\mu}-(1/e)\partial_{\mu}\chi_2$ we obtain
\begin{eqnarray}
K&=&\frac{1}{2}\partial_{\mu}\chi_1\partial^{\mu}\chi_1+\frac{e^2}{2}\left(\frac{\mu^2}{\lambda^2}+\frac{2\mu}{\lambda}\chi_1+\chi_1^2\right)B_{\mu}B^{\mu}
-\frac{1}{4}(\partial_{\mu}B_{\nu}-\partial_{\nu}B_{\mu})(\partial^{\mu}B^{\nu}-\partial^{\nu}B^{\mu}),
\nonumber\\
V(\phi)&=&-\frac{\mu^4}{4\lambda^2}+\mu^2\chi_1^2+\mu\lambda\chi_1^3+\frac{1}{4}\lambda^2\chi_1^4.
\label{M9}
\end{eqnarray}
There is again a Higgs boson field $\chi_1$ with $m^2=+2\mu^2$. However, the field $\chi_2$ has disappeared completely. Instead the vector boson now has a {\bf nonzero mass} given by $m=e\mu/\lambda$. Since a massive gauge boson has three degrees of freedom (two transverse and one longitudinal) while a massless gauge boson such as the photon only has two transverse degrees of freedom, the would-be massless Goldstone boson is absorbed into the now massive gauge boson to provide its needed longitudinal degree of freedom. Hence a massless Goldstone boson and a massless gauge boson are replaced by one massive gauge boson, with two long-range interactions being replaced by one short range interaction. This is known as the Higgs mechanism (1964) though it was initially found by Anderson (1958) in his study of the Meissner effect in superconductivity. The remaining fourth of the original four degrees of freedom becomes the massive Higgs boson, and its presence is an indicator that the Higgs mechanism has taken place. However, the cosmological constant problem remains as severe as before.

\section{The Physics Behind Broken Symmetry}

To discuss broken symmetry in quantum field theory it is convenient to introduce local sources. In the Gell-Mann-Low adiabatic switching procedure one introduces a quantum-mechanical Lagrangian density  $\hat{{{\cal L}}}_0$ of interest, switches on a local source $J(x)$ for some field $\hat{\phi}(x)$ at time $t=-\infty$ and switches it off at $t=+\infty$. While the source is active the Lagrangian of the theory is given by $\hat{{{\cal L}}}_J=\hat{{{\cal L}}}_0+J(x)\hat{\phi}(x)$. Before the source is switched on the system is in the ground state $|\Omega_0^-\rangle$ of the Hamiltonian density $\hat{H}_0$ associate with $\hat{{{\cal L}}}_0$, and after the source is switched off the system is in the state $|\Omega_0^+\rangle$. While $|\Omega_0^-\rangle$ and $|\Omega_0^-\rangle$ are both eigenstates of  $\hat{H}_0$, they differ by a phase, a phase that is fixed by $J(x)$ according to 
\begin{eqnarray}
\langle \Omega_0^+|\Omega_0^-\rangle |_J=\langle \Omega_J|T\exp\left[i\int d^4x(\hat{{{\cal L}}}_0+J(x)\hat{\phi}(x))\right]|\Omega_J\rangle=e^{iW(J)},
\label{M10}
\end{eqnarray}
with this expression serving to define the functional $W(J)$. As introduced, $W(J)$ serves as the generator of the connected $J=0$ theory Green's functions $G^{n}_0(x_1,...,x_n)=\langle\Omega_0|T[\phi(x_1)...\phi(x_n)]|\Omega_0\rangle$ according to
\begin{eqnarray}
W(J)=\sum_n\frac{1}{n!}\int d^4x_1...d^4x_nG^{n}_0(x_1,...,x_n)J(x_1)...J(x_n).
\label{M11}
\end{eqnarray}
On Fourier transforming the Green's functions, we can expand $W(J)$  about the point where all momenta vanish, to obtain
\begin{eqnarray}
W(J)=\int d^4x\left[-\epsilon(J)+\frac{1}{2}Z(J)\partial_{\mu}J\partial^{\mu}J+....\right],
\label{M12}
\end{eqnarray}
with the physical significance of $\epsilon(J)$ (which is spacetime independent if $J$ is) being that it is the energy density difference
\begin{eqnarray}
\epsilon(J)=\langle \Omega_J|\hat{H}_J|\Omega_J\rangle-\langle \Omega_0|\hat{H}_0|\Omega_0\rangle.
\label{M13}
\end{eqnarray}

Given $W(J)$, via functional variation we can construct the so-called classical (c-number) field $\phi_C(x)$ 
\begin{eqnarray}
\phi_C(x)=\frac{\delta W}{\delta J(x)}=\frac{\langle \Omega^+|\hat{\phi}(x)|\Omega^-\rangle}{\langle \Omega^+|\Omega^-\rangle}\bigg{|}_J
\label{M14}
\end{eqnarray}
and the effective action functional
\begin{eqnarray}
\Gamma(\phi_C)=W(J)-\int d^4x J(x)\phi_C(x)=\sum_n\frac{1}{n!}\int d^4x_1...d^4x_n\Gamma^{n}_0(x_1,...,x_n)\phi_C(x_1)...\phi_C(x_n),
\label{M15}
\end{eqnarray}
where the $\Gamma^{n}_0(x_1,...,x_n)$ are the one-particle-irreducible, $\phi_C=0$, Green's functions of $\hat{\phi}(x)$.  Functional variation of $\Gamma(\phi_C)$ then yields
\begin{eqnarray}
\frac{\delta \Gamma(\phi_C)}{\delta \phi_C}=\frac{\delta W}{\delta J }\frac{\delta J}{\delta \phi_C}-J-\frac{\delta J}{\delta \phi_C} \phi_C=-J,
\label{M16}
\end{eqnarray}
to relate $\delta \Gamma(\phi_C)/\delta \phi_C$ back to the source $J$.

On expanding around the point where all external momenta vanish, we can write $\Gamma(\phi_C)$ as
\begin{eqnarray}
\Gamma(\phi_C)=\int d^4x\left[-V(\phi_C)+\frac{1}{2}Z(\phi_C)\partial_{\mu}\phi_C\partial^{\mu}\phi_C+....\right].
\label{M17}
\end{eqnarray}
The quantity 
\begin{eqnarray}
V(\phi_C)=\sum_n\frac{1}{n!}\Gamma^{n}_0(q_i=0)\phi_C^n
\label{M18}
\end{eqnarray}
is known as the effective potential (which is spacetime independent if $\phi_C$ is), while the $Z(\phi_C)$ term  serves as the  kinetic energy of $\phi_C$. The $\Gamma^{n}_0(q_i=0)$ Green's functions can contain two kinds of contributions, tree approximation graphs that involve vertex interactions but no loops, and radiative correction graphs that do contain loops.  For constant $\phi_C$ and $J$ the effective potential is related to the source via $dV/d\phi_C=J$, so that $J$ does indeed break any symmetry that $V(\phi_C)$ might possess. The significance of $V(\phi_C)$ is that when $J$ is zero, we can write $V(\phi_C)$ as 
\begin{eqnarray}
V(\phi_C)=\langle S|\hat{H}_0|S\rangle-\langle N|\hat{H}_0|N\rangle,
\label{M19}
\end{eqnarray}
where $|S\rangle$ and $|N\rangle$ are spontaneously broken and normal vacua in which $\langle S|\hat{\phi}|S\rangle$ is nonzero and $\langle N|\hat{\phi}|N\rangle$ is zero. In the analyses of classical potentials such as $V(\phi)=\lambda^2\phi^4/4-\mu^2\phi^2/2$ and classical kinetic energies such as $K=(1/2)\partial_{\mu}\phi\partial^{\mu}\phi$ presented above, the classical field $\phi$ represented $\phi_C$, the potential $V(\phi)$ represented $V(\phi_C)$, the kinetic energy represented $(1/2)Z(\phi_C)\partial_{\mu}\phi_C\partial^{\mu}\phi_C$, and in  the $\Gamma^{n}_0(q_i=0)$ Green's functions only tree approximation graphs were included. In this way the search for non-trivial minima of $V(\phi)$ is actually a search for states $|S\rangle$ in which $V(\phi_C)=\langle S|\hat{H}_0|S\rangle-\langle N|\hat{H}_0|N\rangle$ would be negative. Thus while the analyses presented above in Sec. IV looked to be classical they actually had a quantum-mechanical underpinning with the classical field being a c-number vacuum matrix element of a q-number quantum field. It is in this way that the classical analyses presented above are to be understood.

To understand the nature of a broken symmetry vacuum it is instructive to reconsider $\epsilon(J)$. It is associated with a system $\hat{H}_0$ to which an external field has been added. This external field breaks the symmetry by hand at the level of the Lagrangian since an effective potential such as  $\epsilon(J)=V(\phi)-J\phi=\lambda^2\phi^4/4-\mu^2\phi^2/2-J\phi$ would be lopsided with one of its minima lower than the other, and would not have any $\phi \rightarrow-\phi$ symmetry. Such a situation is analogous to that found in a ferromagnet. In the presence of an external magnetic field (cf. $J$) all the spins line up in the direction of the magnetic field. If one is above the critical temperature, then if one removes the magnetic field the spins flop back into a configuration in which the net magnetization is zero. However, if one is below the critical point, the spins stay aligned and remember the direction of the magnetic field after it has been removed (hysteresis). Moreover, if the magnetic field is taken to point in some other direction, below the critical point the spins will remember that direction instead, and will remain aligned in that particular direction after the magnetic field is removed. For a spherically symmetric ferromagnetic system below the critical point one can thus align the magnetization at any angle $\theta$ over a full $0$ to $2\pi$ range, and have it remain aligned after the magnetic field is removed.

Given all the different orientations of the magnetization that are possible below the critical point, we need to determine in what way we can  distinguish  them. So consider a single spin pointing in the $z$-direction with spin up, and a second spin pointing at an angle $\theta$ corresponding to a rotation through an angle $\theta$ around the $y$ axis. For these two states the overlap is given by 
\begin{eqnarray}
\langle 0|\theta \rangle=(1,0)e^{-i\theta\sigma_y/2}\left(\matrix{1\cr 0\cr}\right)=(1,0)\left(\matrix{\cos(\theta/2)& \sin(\theta/2)\cr -\sin(\theta/2)&\cos(\theta/2)\cr}\right)\left(\matrix{1\cr 0\cr}\right)=\cos(\theta/2),
\label{M20}
\end{eqnarray}
with the overlap being nonzero and with the two states thus necessarily being in the same Hilbert space. Suppose we now take an $N$-dimensional ensemble of these same sets of spin states and evaluate the overlap of the state with all spins pointing in the z-direction and the state with all spins pointing at an angle $\theta$. This gives the overlap 
\begin{eqnarray}
\langle 0, N|\theta , N\rangle=\cos^N(\theta/2).
\label{M21}
\end{eqnarray}
In the limit in which $N$ goes to infinity this overlap goes to zero. With this also being true of excitations built out of these states, the two states are now in different Hilbert spaces. Thus broken symmetry corresponds to the existence of different, inequivalent  vacua, and even though the various vacua  all have the same energy (i.e. degenerate vacua), the vacua are all in different Hilbert spaces. The quantum Hilbert spaces associated with the various minima of the effective potential (i.e. the differing states $|\Omega\rangle$ in which the vacuum expectation values $\langle \Omega|\hat{\phi}(x)|\Omega\rangle$ are evaluated) become distinct in the limit of an infinite number of degrees of freedom, even though they would not be distinct should $N$ be finite. Broken symmetry is thus not only intrinsically quantum-mechanical, it is intrinsically a many-body effect associated with an infinite number of degrees of freedom.

Whether or not a Hamiltonian $\hat{H}_0$ possesses such a set of degenerate vacua is a property of $\hat{H}_0$ itself. It is not a property of the external field $J$. The role of the external field is solely to pick one of the vacua, so that the system will then remain in that particular vacuum after the external field is removed. Whether or not the system is actually able to remember the direction of the external field after it has been removed is a property of the system itself and not of the external field.

To underscore the need for an infinite number of degrees of freedom, consider a system with a finite number of degrees of freedom such as the one-dimensional, one-body, quantum-mechanical system with potential $V(x)=\lambda^2 x^4/4-\mu^2x^4/2$ and Hamiltonian $H=-(1/2m)\partial^2/\partial x^2+V(x)$. Like the field-theoretic $V(\phi)=\lambda^2\phi^4/4-\mu^2\phi^2/2$, the potential $V(x)$ has a double-well structure, with minima  at $x=\pm \mu/\lambda$. However the eigenstates of the Hamiltonian cannot be localized around either of these two minima. Rather, since the Hamiltonian is symmetric under $x \rightarrow -x$, its eigenstates can only be even functions or odd functions of $x$, and must thus must take support in both of the two wells. Wave functions localized to either of the two wells are in the same Hilbert space, as are then linear superpositions of them, with it being the linear combinations that are the eigenstates. Thus with a finite number of degrees of freedom, wave functions localized around the two minima are in the same Hilbert space. It is only with an infinite number of degrees of freedom that one could get inequivalent Hilbert spaces.

Now if the role of $J$ is only to pick a vacuum and not to make the chosen state actually be a vacuum,  we need to inquire what is there about the $J$ dependence of the theory that might tell us whether or not we do finish up in a degenerate vacuum when we let $J$ go to zero. The answer to this question  is contained in $\epsilon(J)$, with $\epsilon(J)$ needing to be a multiple-valued function of $J$, with $\langle \Omega_J|\hat{\phi}|\Omega_J\rangle$ vanishing on one branch of $\epsilon(J)$ in the limit in which $J$ goes to zero, while not vanishing on some other one.  As a complex function of $J$ the function $\epsilon(J)$ has to have one or more branch points in the complex $J$ plane, and thus have some inequivalent determinations as $J$ goes to zero. These different determinations correspond to different phases, with it being the existence of such inequivalent determinations that is the hallmark of phase transitions.

To appreciate the point consider the two-dimensional Ising model of a ferromagnet in the presence of an external magnetic field $B$ at temperature $T$. In the mean-field approximation the free energy per particle is given by 
\begin{eqnarray}
\frac{F(B,T)}{N}=\frac{1}{2}kT_CM^2-kT\ln 2-kT\ln\left[\cosh\left(\frac{T_CM}{T}+\frac{B}{kT}\right)\right],
\label{M22}
\end{eqnarray}
where $T_C$ is the critical temperature and $M$ is the magnetization. At the minimum where $dF/dM=0$  
the magnetization obeys
\begin{eqnarray}
M=\tanh\left(\frac{T_CM}{T}+\frac{B}{kT}\right).
\label{M23}
\end{eqnarray}
When $T$ is greater than $T_C$ the magnetization can only be nonzero if $B$ is nonzero. However, if $T$ is less than $T_C$ one can have a nonzero $M$ even if $B$ is zero, and not only that, for every non-trivial $M$ there is another solution with $-M$. Since $\cosh(T_CM/T)$ is an even function of $M$, solutions of either sign for $M$ have the same  free energy. Symmetry breaking is thus associated with a degenerate vacuum energy. If we take $B$ to be complex, set $B=B_R+iB_I$, and set $\alpha=T_CM/T+B_R/kT$, $\beta=B_I/kT$, then when $B$ is nonzero we can set $\cosh(\alpha+i\beta)=\cosh\alpha\cos\beta+i\sinh\alpha\sin\beta$. With the logarithm term in the free energy having branch points in the complex $B$ plane whenever $\cosh(\alpha+i\beta)=0$, we see that branch points occur when $\alpha=0$, $\beta=\pi/2, 3\pi/2, 5\pi/2,...$. Thus as required, the free energy is a multiple-valued-function in the complex $B$ plane, with branch points on the imaginary $B$ axis. While $M=\tanh(T_CM/T)$ only has two real solutions for any given $T<T_C$, it has an infinite number of pure imaginary solutions, and these are reflected in the locations of the branch points of $F(B,T)$.

A second example of multiple-valuedness  may be found in the double-well potential  $V(\phi)=\lambda^2\phi^4/4-\mu^2\phi^2/2$ given in Sec. IVA in the presence of a constant source $J$. Solutions to the theory are constrained to obey 
\begin{eqnarray}
\frac{dV(\phi)}{d \phi}=\lambda^2\phi^3-\mu^2\phi=J,
\label{M24}
\end{eqnarray}
and are of the form
\begin{eqnarray}
\phi=i^{1/3}(p+iq)^{1/3}+[i^{1/3}(p+iq)^{1/3}]^*,
\label{M25}
\end{eqnarray}
where
\begin{eqnarray}
p=\left(\frac{\mu^6}{27\lambda^6}-\frac{J^2}{4\lambda^2}\right)^{1/2},\qquad q=-\frac{J}{2\lambda^2}.
\label{M26}
\end{eqnarray}
If we set $i^{1/3}=\exp(-i\pi/2), \exp(i\pi/6), \exp(5i\pi/6)$, then when $J=0$, the solutions are given by $\phi_1=0$, $\phi_2=\mu/\lambda$, $\phi_3=-\mu/\lambda$, just as found in Sec IVA. 

However, suppose instead we fix $i^{1/3}=\exp(-i\pi/2)$, and treat $\phi$ as a multiple-valued function of $J$. Then, because of the cube root in the $(p+ iq)^{1/3}$ term, as we set $J$ to zero we obtain three determinations of $p^{1/3}$, viz. $p_1=\mu/\lambda\surd{3}$, $p_2=\exp(2\pi i/3)\mu/\lambda\surd{3}$, $p_3=\exp(4\pi i/3)\mu/\lambda\surd{3}$. With $J=0$, these determinations then precisely give the previous $\phi_1=0$, $\phi_2=\mu/\lambda$, $\phi_3=-\mu/\lambda$ solutions. With this multiple-valuedness then propagating to $V(\phi)$ and $\epsilon(J)=V(\phi)-J\phi$ when they are evaluated in these three solutions, i.e. when we set $dV(\phi)/d\phi=J$, $d\epsilon(J)/dJ=-\phi$ and obtain 
\begin{eqnarray}
\epsilon(J)&=&-\frac{3\lambda^2}{4}\phi^4+\frac{\mu^2}{2}\phi^2
\nonumber\\
&=&-\frac{\mu^2}{4}\phi^2-\frac{3}{4}\phi J
\nonumber\\
&=&-\frac{3J}{4}\left[i^{1/3}(p+iq)^{1/3}+{\rm c.~c.}\right]-\frac{\mu^2}{4}\left[i^{2/3}(p+iq)^{2/3}+{\rm c.~c.}\right]-\frac{\mu^4}{6\lambda^2},
\label{M27}
\end{eqnarray}
we see that  in any solution $\epsilon(J)$ is indeed a multiple-valued function of $J$, and see that from any one solution we can derive the others by analytic continuation, with the limit $J \rightarrow 0$ having multiple determinations. 

While we need to use the effective potential $V(\phi_C)$ to explore symmetry breaking by an elementary scalar field, it turns out that even though $W(J)$ was only introduced as an intermediate step in order to get to $\Gamma(\phi_C)$, for  symmetry breaking by fermion $\bar{\psi}\psi$ bilinears we need to use $W(m)$ directly, where
\begin{eqnarray}
e^{iW(m)}=\langle \Omega_0^+|\Omega_0^-\rangle=\langle \Omega_m|T\exp\left[i\int d^4x(\hat{{{\cal L}}}_0-m(x){\bar{\psi}}(x)\psi(x))\right]|\Omega_m\rangle.
\label{M28}
\end{eqnarray}
Specifically, consider the chiral symmetric Nambu-Jona-Lasinio (NJL) action with no fermion mass term of the form
\begin{equation}
I_{\rm NJL}=\int d^4x \left[i\bar{\psi}\gamma^{\mu}\partial_{\mu}\psi-\frac{g}{2}[\bar{\psi}\psi]^2-\frac{g}{2}[\bar{\psi}i\gamma_5\psi]^2\right].
\label{M29}
\end{equation}
As such it is a relativistic generalization of the BCS model. In the mean field, Hartree-Fock  approximation one introduces a trial wave function parameter $m$ that is not in the original action, and then decomposes  the action into two pieces,  a mean-field piece and a residual-interaction piece according to:
\begin{eqnarray}
I_{\rm NJL}&=&I_{\rm MF}+I_{\rm RI}
\nonumber\\
&=&\int d^4x \left[i \bar{\psi}\gamma^{\mu}\partial_{\mu}\psi-m\bar{\psi}\psi +\frac{m^2}{2g}\right]
+\int d^4x \left[-\frac{g}{2}\left(\bar{\psi}\psi-\frac{m}{g}\right)^2-\frac{g}{2}\left(\bar{\psi}i\gamma_5\psi\right)^2\right].
\label{M30}
\end{eqnarray}
One then tries to show that in the mean-field sector nonzero $m$ is energetically favored. If we define $\hat{{{\cal L}}}_0=i \bar{\psi}\gamma^{\mu}\partial_{\mu}\psi+m^2/2g$  and define $\hat{{{\cal L}}}_m=\hat{{{\cal L}}}_0-m(x)\bar{\psi}(x)\psi(x)$, then in the mean-field sector we need to show that the $\epsilon(m)$ associated with $\hat{{{\cal L}}}_m$ is less than $\epsilon(m=0)$ and thus energetically favored. Now previously we had identified $\epsilon(m)$ as $\epsilon(m)=\langle \Omega_m|\hat{H}_m|\Omega_m\rangle-\langle \Omega_0|\hat{H}_0|\Omega_0\rangle$, where $\hat{H}_m=\hat{H}_0+m\bar{\psi}\psi$. If the state $|\Omega_0\rangle$  is one in which $\langle \Omega_0|\bar{\psi}\psi|\Omega_0\rangle$ is zero, we can write $\epsilon(m)$ as 
\begin{eqnarray}
\epsilon(m)=\langle \Omega_m|\hat{H}_m|\Omega_m\rangle-\langle \Omega_0|\hat{H}_m|\Omega_0\rangle,
\label{M31}
\end{eqnarray}
to thus enable us to compare two candidate vacua for $\hat{H}_m$, with a view to determining whether a vacuum with nonzero $\langle \Omega_m|\bar{\psi}\psi|\Omega_m\rangle$ is energetically favored. In terms of Green's functions $\epsilon(m)$ can be evaluated via
\begin{eqnarray}
\epsilon(m)=\sum_n\frac{1}{n!}G^{n}_0(q_i=0)m^n,
\label{M32}
\end{eqnarray}
where the $G^{n}_0(q_i=0)$ are the connected $\bar{\psi}\psi$ Green's functions as evaluated in the theory in which $m=0$, viz. that associated with $\hat{H}_0$. Unlike the elementary scalar field $V(\phi_C)$, $\epsilon(m)$ has no tree approximation contribution and is entirely generated by radiative loops.

Thus while we need to use $V(\phi_C)$ to explore symmetry breaking by an elementary scalar field with $V(\phi_C)$ being the energy density difference between different candidate vacua of $H_0$, for symmetry breaking by a fermion composite we use $\epsilon(m)$ instead, with $\epsilon(m)$ being the energy density difference between different candidate vacua of $H_m$. We shall explore this issue in more detail below, while noting now that in distinguishing between an elementary (i.e. ``god given") Higgs boson and a composite one there are even differences in setting up the formalism in the first place.

\section{What exactly is the Higgs field?}

Given that the existence of the Higgs boson has now been confirmed, we need to ask what exactly the field $\phi$ represents. There are two possibilities. It is either a q-number elementary field $\phi$ that appears in the fundamental  $SU(3)\times SU(2)\times U(1)$ Lagrangian (to thereby be on an equal footing with the fundamental quarks, leptons and gauge bosons), or it is generated as a dynamical bound state, with the field in a dynamically induced Higgs potential then being the c-number matrix element $\langle \Omega|\bar{\psi}\psi|\Omega\rangle$,  a dynamical bilinear fermion condensate. The Mexican Hat potential is thus either part of the fundamental Lagrangian or it is generated by  dynamics. If the Higgs field is elementary, then while the potential $V(\phi)=\lambda^2\phi^4/4-\mu^2\phi^2/2$ would be its full quantum-mechanical potential, the discussion given earlier of the minima of the potential would correspond to a c-number tree approximation analysis with the $\phi$ that appeared there being the c-number $\langle \Omega|\hat{\phi}|\Omega\rangle$. However, in the fermion condensate case there is no tree approximation, with the theory being given by radiative loop diagrams alone. To see how to generate the Mexican Hat potential in this case we consider the Nambu-Jona-Lasinio (NJL) four-fermion model.

\subsection{Nambu-Jona-Lasinio Chiral Model as a Mean-Field Theory}

The NJL model is a chirally-symmetric four-fermion model of interacting massless fermions with action $I_{\rm NJL}$ as given above. In the mean field, Hartree-Fock  approximation one introduces a trial wave function parameter $m$ that is not in the original action, and then decomposes  the $I_{\rm NJL}$ action into two pieces,  a mean field piece and a residual interaction according to $I_{\rm NJL}=I_{\rm MF}+I_{\rm RI}$, where
\begin{eqnarray}
I_{\rm MF}&=&\int d^4x \left[i \bar{\psi}\gamma^{\mu}\partial_{\mu}\psi-m\bar{\psi}\psi +\frac{m^2}{2g}\right],
\nonumber\\
I_{\rm RI}&=&\int d^4x \left[-\frac{g}{2}\left(\bar{\psi}\psi-\frac{m}{g}\right)^2-\frac{g}{2}\left(\bar{\psi}i\gamma_5\psi\right)^2\right].
\label{M33}
\end{eqnarray}
In the mean field approximation one sets 
\begin{eqnarray}
&&\langle S|\left[\bar{\psi}\psi-\frac{m}{g}\right]^2|S\rangle=\langle S|\left[\bar{\psi}\psi-\frac{m}{g}\right]|S\rangle^2=0,
\nonumber\\
&&\langle S|\bar{\psi}\psi|S\rangle=\frac{m}{g},~~~\langle S|\bar{\psi}i\gamma_5\psi|S\rangle=0,
\label{M34}
\end{eqnarray}
to thus give the residual interaction energy density a zero vacuum expectation value in the state $|S\rangle$. In the mean-field approximation the physical mass $M$ is the value of $m$ that satisfies that satisfies $\langle S|\bar{\psi}\psi|S\rangle=m/g$. The one loop contribution of the fermionic negative energy Dirac sea  to the quantity $\langle S|\bar{\psi}\psi|S\rangle$ yields the gap equation
\begin{equation}
-\frac{M\Lambda^2}{4\pi^2}+\frac{M^3}{4\pi^2}{\rm ln}\left(\frac{\Lambda^2}{M^2}\right)=\frac{M}{g},
\label{M35}
\end{equation}
where $\Lambda$ is an ultraviolet cut-off, as needed since the NJL model is not renormalizable.

\begin{figure}[htpb]
\begin{center}
\includegraphics[width=3.3in,height=1.3in]{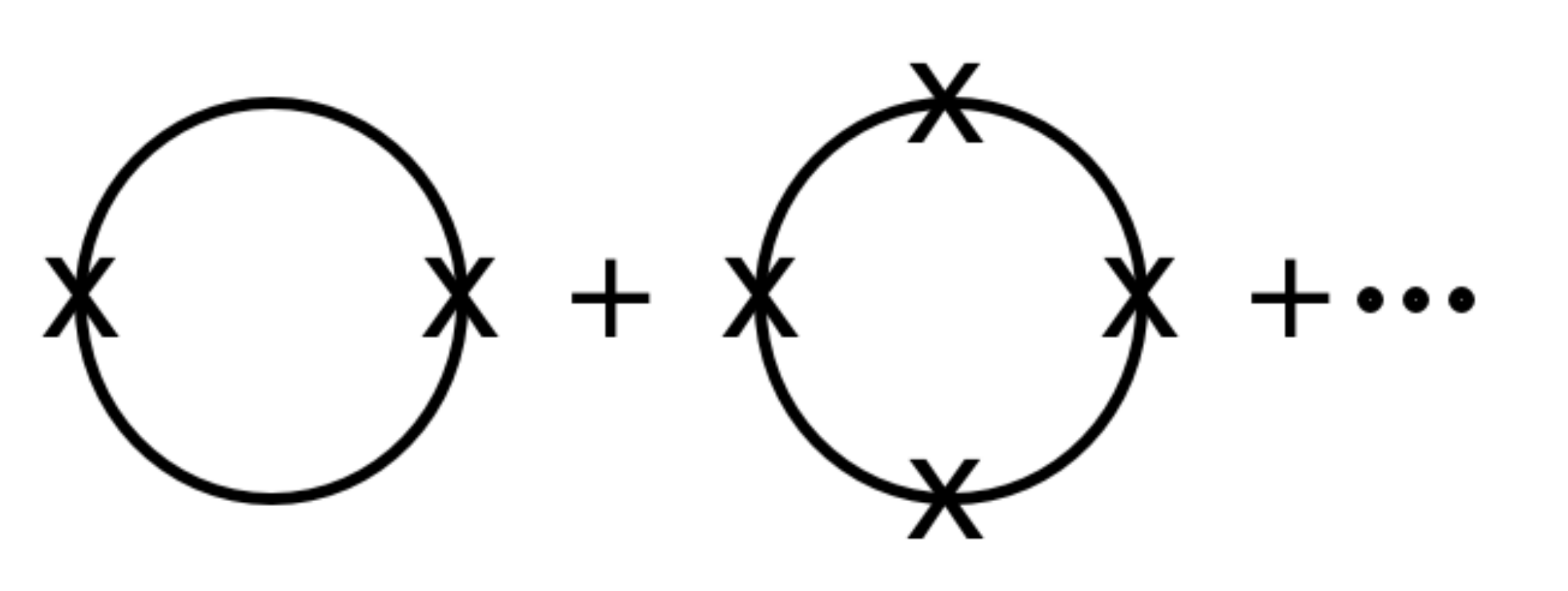}
\caption{Vacuum energy density $\epsilon(m)$ via an infinite summation of massless graphs with zero-momentum point $m\bar{\psi}\psi$  insertions.}
\end{center}
\end{figure}

Given this gap equation, for $I_{\rm MF}$ we can calculate the one loop mean-field vacuum energy density $\epsilon(m)$ as a function of a constant $m$ via the graphs of Fig. (2). This yields:
\begin{eqnarray}
\epsilon(m)&=&\sum_n\frac{1}{n!}G^{n}_0(q_i=0)m^n-\frac{m^2}{2g}
\nonumber\\
&=&i\int \frac{d^4p}{(2\pi)^4}{\rm Tr~ln}\left[\frac{\gamma^{\mu}p_{\mu}-m+i\epsilon}{\gamma^{\nu}p_{  \nu}+i\epsilon}\right]-\frac{m^2}{2g}
\nonumber\\
&=&\frac{m^4}{16\pi^2}{\rm ln}\left(\frac{\Lambda^2}{m^2}\right)-\frac{m^2M^2}{8\pi^2}{\rm ln}\left(\frac{\Lambda^2}{M^2}\right)+\frac{m^4}{32\pi^2}.
\label{M36}
\end{eqnarray}
While the energy $i\int d^4p/(2\pi)^4{\rm Tr~ln}[\gamma^{\mu}p_{\mu}-m]$ has quartic, quadratic and logarithmically divergent pieces, the subtraction of the massless vacuum energy $i\int d^4p/(2\pi)^4{\rm Tr~ln} [\gamma^{\mu}p_{\mu}]$ removes the quartic divergence, with the subtraction of the self-consistent induced mean field term $m^2/2g$ then leaving $\epsilon(m)$ only logarithmically divergent. We recognize the resulting logarithmically divergent  $\epsilon(m)$ as having a local maximum at $m=0$, and a global minimum at $m=M$ where $M$ itself is finite.  We thus induce none other than a dynamical double-well Mexican Hat potential, and identify $M$ as the matrix element of a fermion bilinear according to $M/g=\langle S|\bar{\psi}\psi|S\rangle$.

If instead of looking at matrix elements in the translationally-invariant vacuum $|S\rangle$ we instead look at matrix elements in coherent states $|C\rangle$  where  $m(x)=\langle C|\bar{\psi}(x)\psi(x)|C\rangle$ is now spacetime dependent, we then find \cite{Eguchi1974,Mannheim1976} that the resulting mean field effective action has a logarithmically divergent part  of the form
\begin{eqnarray}
I_{\rm EFF}=\int \frac{d^4x}{8\pi^2}{\rm ln}\left(\frac{\Lambda^2}{M^2}\right)\left[
\frac{1}{2}\partial_{\mu}m(x)\partial^{\mu}m(x)+m^2(x)M^2-\frac{1}{2}m^4(x)\right].
\label{M37}
\end{eqnarray}
Here the kinetic energy term is the analog of the $Z(J)$ term given earlier in the expansion of $W(J)$ around the point where all momenta vanish. In terms of the quantity $ \Pi_{\rm S}(q^2,M)$ to be introduced below, $Z(M)$ is given by $\partial \Pi_{\rm S}(q^2, M)/\partial q^2|_{q^2=0}$.  If we introduce a coupling $g_{\rm A}\bar{\psi}\gamma_{\mu}\gamma_5A^{\mu}_{5}\psi $ to an axial gauge field $A^{\mu}_{5}(x)$, on setting $\phi=\bar{\psi}(1+\gamma_5)\psi$ the effective action becomes 
\begin{eqnarray}
I_{\rm EFF}&=&\int \frac{d^4x}{8\pi^2}{\rm ln}\left(\frac{\Lambda^2}{M^2}\right)\bigg{[}
\frac{1}{2}|(\partial_{\mu}-2ig_{\rm A}A_{\mu 5})\phi(x)|^2+|\phi(x)|^2M^2
-\frac{1}{2}|\phi(x)|^4-\frac{g_{\rm A}^2}{6}F_{\mu\nu 5}F^{\mu\nu 5}\bigg{]}.
\nonumber\\
\label{M38}
\end{eqnarray}
We recognize this action as a double-well Ginzburg-Landau type Higgs Lagrangian, only now generated dynamically. We thus generalize to the relativistic chiral case Gorkov's derivation of the Ginzburg-Landau order parameter action starting from the BCS four-fermion theory. In the $I_{\rm EFF}$ effective action associated with the NJL model there is a double-well Higgs potential, but since $m(x)=\langle C|\bar{\psi}(x)\psi(x)|C\rangle$ is a c-number, $m(x)$ does not itself represent a q-number scalar field. And not only that, unlike in the elementary Higgs case, the second derivative of $V(m)$ at the minimum where $m=M$ is not the mass of a q-number Higgs boson. Rather, as we now show, the q-number fields are to be found as collective modes generated by the residual interaction, and it is the residual interaction that will fix their masses. Moreover, as we will see in Sec. VII, when we dress the point NJL vertices in Figs. (2) and (3), the dynamical Higgs boson will move above the threshold in the fermion-antifermion scattering amplitude, become unstable,  and acquire a width. This width has the potential to discriminate between an elementary Higgs boson and a dynamical one.

\subsection{The Collective Goldstone and Higgs Modes}

To find the collective modes we calculate the scalar and pseudoscalar sector Green's functions $\Pi_{\rm S}(x)=\langle \Omega|T(\bar{\psi}(x)\psi(x)\bar{\psi}(0)\psi(0))|\Omega\rangle$, $\Pi_{\rm P}(x)=\langle \Omega|T(\bar{\psi}(x)i\gamma_5\psi(x) \bar{\psi}(0)i\gamma_5\psi(0))|\Omega\rangle$, as is appropriate to a chiral-invariant theory. If first we take the fermion to be massless (i.e. setting $|\Omega\rangle=|N\rangle$ where $\langle N|\bar{\psi}\psi|N\rangle=0$), to one loop order in the four-fermion residual interaction we obtain

\begin{equation}
\Pi_{\rm S}(q^2, M=0)=\Pi_{\rm P}(q^2, M=0)=-\frac{1}{8\pi^2}\left( 2\Lambda^2+q^2{\rm ln}\left(\frac{\Lambda^2}{-q^2}\right)+q^2\right).
\label{M39}
\end{equation}
The scattering matrices in the two channels are given by 
\begin{equation}
T_{\rm S}(q^2, M=0)=\frac{1}{g^{-1}-\Pi_{\rm S}(q^2, M=0)},~~~T_{\rm P}(q^2, M=0)=\frac{1}{g^{-1}-\Pi_{\rm P}(q^2, M=0)},
\label{M40}
\end{equation}
and with $g^{-1}$ given by the gap equation, near $q^2=-2M^2$ both scattering matrices behave as
\begin{equation}
T_{\rm S}(q^2, M=0)=T_{\rm P}(q^2, M=0)=\frac{Z^{-1}}{(q^2+2M^2)},~~~~~
Z=\frac{1}{8\pi^2}{\rm ln}\left(\frac{\Lambda^2}{M^2}\right),
\label{M41}
\end{equation}
to give degenerate (i.e. chirally symmetric) scalar and pseudoscalar tachyons at $q^2=-2M^2$ (just like fluctuating around the local maximum in a double-well potential), with $|N\rangle$ thus being unstable.

\begin{figure}[htpb]
\includegraphics[width=3.4in,height=0.8in]{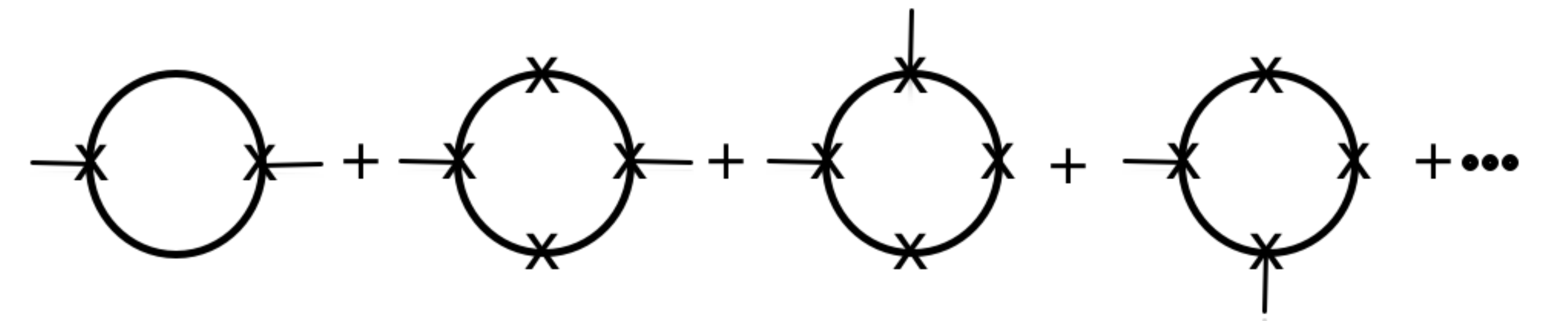}
\caption{$\Pi_{\rm S}(q^2,m)$ developed as an infinite summation of massless graphs, each with two point $m\bar{\psi}\psi$ insertions carrying momentum $q_{\mu}$ (shown as external lines), with all other point $m\bar{\psi}\psi$ insertions carrying zero momentum.}
\end{figure}

However, suppose we now take the fermion to have nonzero mass $M$ (i.e. we set $|\Omega\rangle=|S\rangle$). As per the summation for $\Pi_{\rm S}(q^2,m)$ given in Fig. (3) and its $\Pi_{\rm P}(q^2,m)$ analog,  we obtain
\begin{eqnarray}
\Pi_{\rm P}(q^2, M)&=&-\frac{\Lambda^2}{4\pi^2}
+\frac{M^2}{4\pi^2}{\rm ln}\left(\frac{\Lambda^2}{M^2}\right) 
-\frac{q^2}{8\pi^2}{\rm ln}\left(\frac{\Lambda^2}{M^2}\right) 
-\frac{(q^2-4M^2)}{8\pi^2}
\nonumber\\
&-&\frac{(8M^4-8M^2q^2+q^4)}{8\pi^2 q^2}\left(\frac{-q^2}{4M^2-q^2}\right)^{1/2}{\rm ln}\left(\frac{(4M^2-q^2)^{1/2}+(-q^2)^{1/2}}{(4M^2-q^2)^{1/2}-(-q^2)^{1/2}}\right),
\nonumber\\
\Pi_{\rm S}(q^2, M)&=&-\frac{\Lambda^2}{4\pi^2}
+\frac{M^2}{4\pi^2}{\rm ln}\left(\frac{\Lambda^2}{M^2}\right)
+\frac{(4M^2-q^2)}{8\pi^2}{\rm ln}\left(\frac{\Lambda^2}{M^2}\right) 
+\frac{(4M^2-q^2)}{8\pi^2}
\nonumber\\
&-&\frac{(4M^2-q^2)}{8\pi^2}\left(\frac{4M^2-q^2}{-q^2}\right)^{1/2}{\rm ln}\left(\frac{(4M^2-q^2)^{1/2}+(-q^2)^{1/2}}{(4M^2-q^2)^{1/2}-(-q^2)^{1/2}}\right).
\label{M42}
\end{eqnarray}
Given the form for $g^{-1}$,  we find a dynamical pseudoscalar Goldstone boson bound state at $q^2=0$ and a  dynamical scalar Higgs boson bound state at $q^2=4M^2$ ($=-2\times  M^2({\rm tachyon})$). The two dynamical bound states are not degenerate in mass (spontaneously broken chiral symmetry), and the dynamical Higgs  scalar mass $2M$ is twice the induced mass of the fermion.

\section{What Comes Next?}

The most crucial question for the Higgs boson is determining its underlying nature. Thus we need to find some experimental way to determine whether the Higgs boson is elementary or dynamical. If the Higgs field is elementary, one would have to treat the Higgs Lagrangian as a bona fide quantum field theory with an a priori double-well potential being present in the fundamental Lagrangian itself. The great appeal of a Higgs potential of the specific form $V(\phi)=\lambda^2\phi^4/4-\mu^2\phi^2/2$ is that the radiative corrections that it generates are renormalizable. Thus not only would the massive gauge boson sector of the theory be renormalizable (because of the Higgs mechanism), the Higgs sector itself would be too. Moreover, with an elementary Higgs field electroweak radiative corrections are not only finite, they are straightforwardly calculable.

However, having an elementary Higgs field also has some disturbing consequences.  First we note that the radiative corrections in the Higgs sector itself lead to a quadratically divergent Higgs self energy. While this divergence can be made finite by renormalization, there is no control on the ensuing magnitude that the Higgs mass would take, as it would be as large as the regulator masses. These masses could be at the $10^{16}$ GeV or so grandunified scale (a very large scale if the proton lifetime is to be greater than  its current experimental lower bound) or even at the $10^{19}$ GeV quantum gravitational Planck scale. Such mass values are orders of magnitude larger than the Higgs boson's observed 125 GeV mass. In order to resolve such a disparity (known as the hierarchy problem) it had long been conjectured that there would be a supersymmetry between bosons and fermions, and with fermion and boson loops having opposite signs, a fermionic  superparticle could then cancel  the quadratic divergence in the Higgs self-energy. For it to do so to the degree needed, the fermionic superparticle would need to be close in mass to the Higgs boson itself. However, data from the very same LHC that was used to find the Higgs boson have sharply curtailed the possibility  that there might be any fermionic superparticle in the requisite mass region. Thus the Higgs self-energy problem is open at the present time.

A second concern for an elementary Higgs field is that the group theoretic structure of the $SU(2)\times U(1)$ theory does not at all constrain the Yukawa couplings of the Higgs field to the fundamental fermions. Thus while the Higgs field can give masses to the quarks and leptons, those masses are totally unconstrained, with all the Yukawa coupling constants being free parameters that one has to introduce by hand. While embedding $SU(2)\times U(1)$ in some grandunified theory of the strong, electromagnetic and weak interactions, say, might solve this problem, the issue is open at the present time.

A  third concern for an elementary Higgs field is that, as had been noted earlier, the very minimization of the Higgs potential generates an enormous contribution to the cosmological constant. For a one TeV or so Higgs mass breaking scale one would get a cosmological constant that would be of order $10^{60}$ times larger than the standard Einstein gravitational theory could possibly permit. This problem is very severe, and it also is open at the present time.  

A fourth concern for an elementary Higgs field is an aesthetic, in principle one. At the level of the $SU(3)\times SU(2)\times U(1)$ Lagrangian all fermions and gauge bosons are massless and all coupling constants are dimensionless. The only place where there is a fundamental mass scale is in the $-\mu^2\phi^2/2$ term in the Higgs potential. It would be much more natural  and elegant if this $\mu^2$ scale were to be generated dynamically in a then scale invariant theory that would then possess no fundamental scale at all.

It is thus of interest to note that all of these concerns can be addressed if the Higgs boson is dynamical. Moreover, there already are two working models that we know of in which the symmetry breaking actually is done dynamically, the BCS theory, and the generation of a Goldstone boson pion in QCD. In the BCS theory the symmetry breaking is done by a dynamically generated nonzero fermion bilinear vacuum expectation value $\langle S|\psi\psi|S\rangle$ in a theory that only contains electrons and phonons while not containing any elementary scalar fields whatsoever. For the pion, we note that the QCD local color  $SU(3)$ Lagrangian of quarks and gluons possesses a global chiral flavor symmetry. This chiral symmetry is broken dynamically by QCD dynamics to yield a Goldstone pion. This pion then acquires a mass via the weak interaction since the electroweak $SU(2)\times U(1)$ action breaks the flavor chiral symmetry intrinsically at the level of weak interaction Lagrangian itself, to thereby make the chiral favor symmetry be only an approximate one.  

The generation of a Goldstone pion in QCD  is particularly of interest since unlike the NJL model,  in the QCD case it occurs in a theory that is renormalizable, and in which all coupling constants are dimensionless  and all quark and gluon masses are zero at the level of the Lagrangian.  In such a theory not only is there no fundamental $-\mu^2\phi^2/2$ term, there is not even any elementary $\phi$ at all, and the theory is scale invariant at the level of the Lagrangian.  So let us suppose that dynamical symmetry breaking occurs in some renormalizable Yang-Mills theory of interacting massless fermions and gauge bosons. In such a case, if nonzero gauge boson masses are produced by a dynamical Higgs mechanism, renormalizability would not be impaired since bound state production in a renormalizable theory does not violate renormalizability. Without needing to specify any particular such Yang-Mills theory (i.e. without needing to specify the group structure or the specific matter content of the theory), we note that if the symmetry breaking occurs at all, then since all such theories have no divergences higher than logarithmic, there will be no quadratic divergence associated with any dynamical scalar bound states that might be produced.  Thus with a dynamical Higgs boson there is no Higgs self-energy hierarchy problem at all. 

As regards the couplings of fermions to the Higgs boson, these couplings are given as the residues of the dynamical poles in the requisite channels. Thus they are determined by the theory itself and are not free parameters at all. To see how things work, let us consider the NJL model as a stand-in for a renormalizable field theory. In its $T_{\rm S}$ and $T_{\rm P}$ channels there are scalar and pseudoscalar bound states, and near the respective poles the scattering amplitudes behave as: 
\begin{equation}
T_{\rm S}(q^2, M)=\frac{Z^{-1}_{\rm S}}{(q^2-4M^2)},~~~~T_{\rm P}(q^2, M)=\frac{Z^{-1}_{\rm P}}{q^2},~~~~
Z_{\rm S}=Z_{\rm P}=\frac{1}{8\pi^2}{\rm ln}\left(\frac{\Lambda^2}{M^2}\right).
\label{M43}
\end{equation}
\noindent
We thus identify the Yukawa coupling of the scalar and psuedoscalar bound states to a fermion anti-fermion pair to be $Z_{\rm S}^{-1/2}$ and $Z_{\rm P}^{-1/2}$, where the values of $Z_{\rm S}$ and $Z_{\rm P}$ are not fixed by hand but by the theory itself, with the equality of $Z_{\rm S}$ and $Z_{\rm P}$ that is found reflecting the underlying chiral symmetry of the NJL theory.

As regards the cosmological constant problem, we had noted above that the self-consistent mean-field treatment of the NJL model automatically generated an $m^2/2g$ term in the mean-field action, with this term serving to cancel the quadratic divergence in the vacuum energy. Unlike the situation with an elementary Higgs field where there is no control of the vacuum energy, with dynamical symmetry breaking we see that contributions to the vacuum energy are potentially controllable. Breaking symmetries dynamically thus provides a good starting point to address the cosmological constant problem. 

\begin{figure}[htpb]
\includegraphics[width=3.3in,height=1.3in]{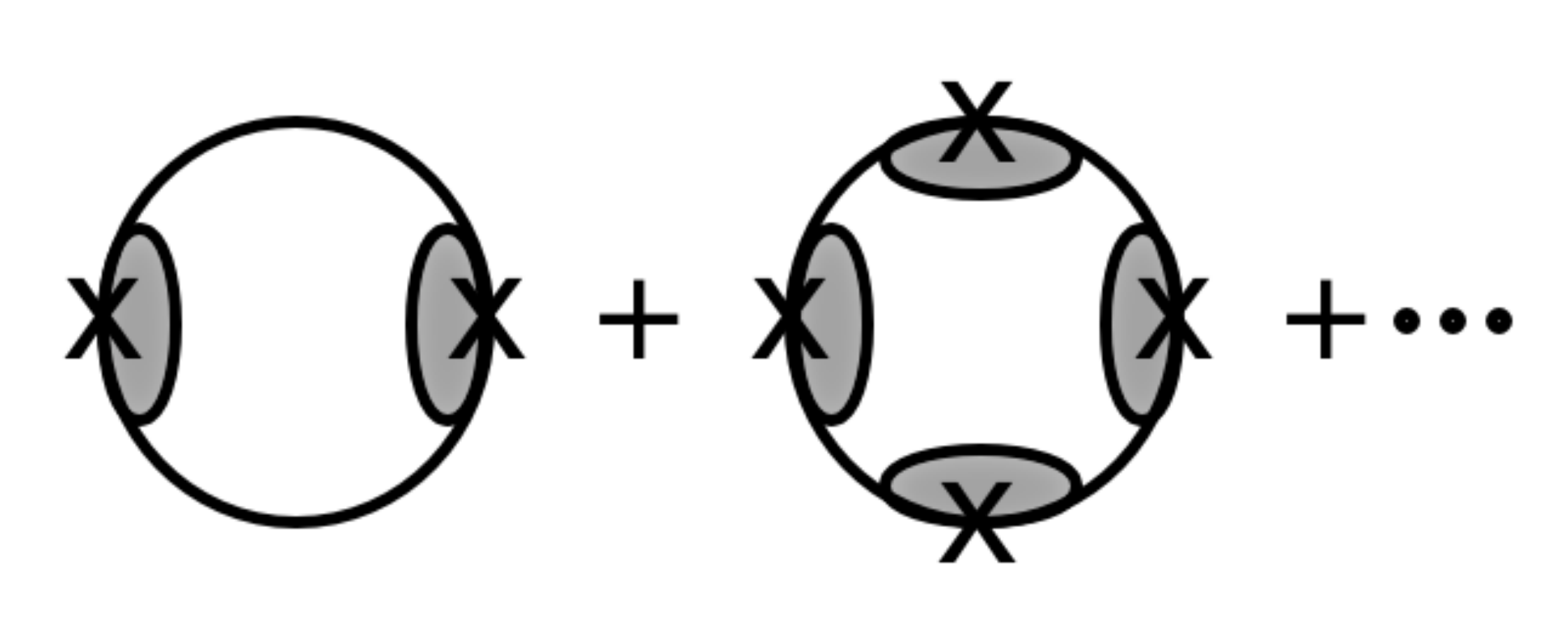}
\caption{Vacuum energy density  $\epsilon(m)$ via an infinite summation of massless graphs with zero-momentum dressed $m\bar{\psi}\psi$ insertions.}
\end{figure}
\begin{figure}[htpb]
\includegraphics[width=3.4in,height=0.8in]{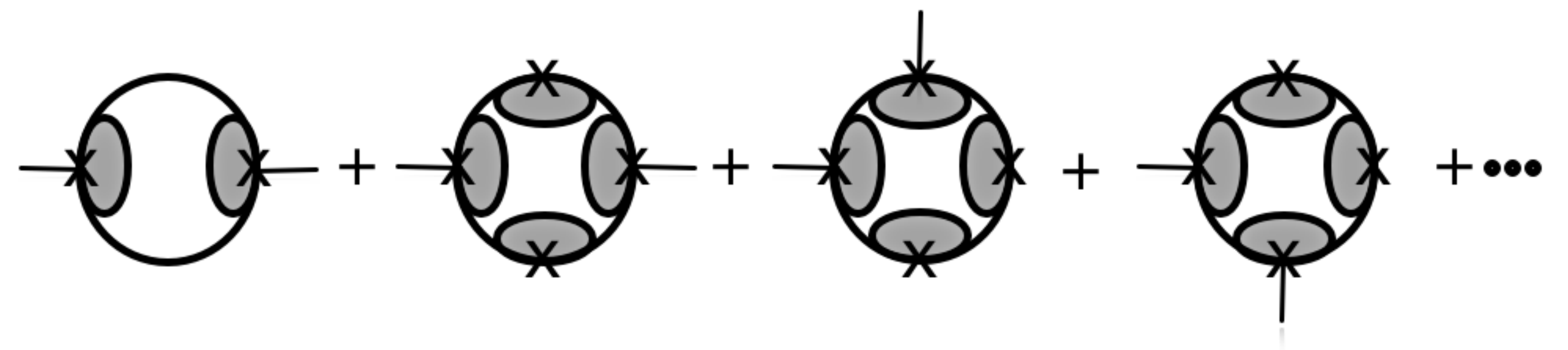}
\caption{$\Pi_{\rm S}(q^2,m)$ developed as an infinite summation of massless graphs, each with two dressed $m\bar{\psi}\psi$  insertions carrying momentum $q_{\mu}$ (shown as external lines), with all other dressed $m\bar{\psi}\psi$  insertions carrying zero momentum.}
\end{figure}

A discussion of how things might work in the renormalizable situation can  be found in the conformal gravity and conformal symmetry studies given in   \cite{Mannheim2012} and \cite{Mannheim2015}. In these studies conformal symmetry is used to address the cosmological constant problem -- with the cosmological constant  being zero if the conformal symmetry is unbroken,  while being constrained in a phenomenologically acceptable way if the conformal symmetry is broken dynamically. In addition, the conformal symmetry is realized via scaling with anomalous dimensions, with the Green's function associated with the insertion of a soft $\bar{\psi}\psi$ into the inverse fermion propagator being found to behave as $\Gamma_{\rm S}(p,p,0)=(p^2)^{-1/2}$. With such a form the dynamical dimension of $\bar{\psi}\psi$ is reduced from three to two, and the dynamical dimension of $[\bar{\psi}\psi]^2$ is reduced from six to a renormalizable four. The four-fermion theory is thus made renormalizable by dressing its point vertices in this particular way, with the point-coupled vertices of the NJL model with $\Gamma_{\rm S}(p,p,0)=1$ being replaced by dressed vertices with $\Gamma_{\rm S}(p,p,0)=(p^2)^{-1/2}$.  With the $\Gamma_{\rm S}(p,p,0)=(p^2)^{-1/2}$ dressing all infinities in the particle physics sector are now under control, and with conformal gravity being a renormalizable (and unitary \cite{Bender2008a,Bender2008b}) theory, all infinities in both particle physics and gravity (and thus the infinite negative energy sea contribution  to the cosmological constant) are now under control.

For our purposes here we note that once one dresses the vertices as exhibited in Figs. (4) and (5), quantities such $\Pi_{\rm S}(q^2,M)$  will develop a non-trivial momentum dependence, so that $\Pi_{\rm S}(q^2,M)$ will then have a discontinuity in the complex $q^2$ plane.  In consequence of this, the pole in  $T_{\rm S}(q^2,M)$ will no longer be at the particle-antiparticle threshold but will instead be above threshold and not be on the real $q^2$ axis \cite{Mannheim2015}. The dynamically induced Higgs boson must thus be an unstable above-threshold resonance and have a nonzero width. It is in this respect that a composite Higgs boson will most differ from an elementary one, with the width of the Higgs boson thus being a diagnostic that could potentially distinguish an elementary Higgs boson from a dynamical one.

However, if the Higgs boson is to be dynamical, the dynamical theory in which it is to be generated would then have to be able to reproduce those aspects of the radiative correction structure associated with an elementary Higgs field theory that have been tested. To this end we note that following Gaussian integration on a dummy scalar field variable $\sigma$ and a dummy vector field variable $V_{\mu}$, the path integral representation of the generator $Z(\bar{\eta},\eta)$ of fermion Green's functions in the NJL theory can be written as:
\begin{eqnarray}
Z(\bar{\eta}, \eta)&=&\int [d\bar{\psi}][d\psi]\exp\left[i\int d^4x \left(i\bar{\psi}\gamma^{\mu}\partial_{\mu}\psi-\frac{g}{2}(\bar{\psi}\psi)^2+\bar{\eta}\psi+\bar{\psi}\eta\right)\right]
\nonumber\\
&=&\int [d\bar{\psi}][d\psi][ d\sigma][dV_{\mu}]\exp\bigg{[}i\int d^4x \bigg{(}i\bar{\psi}\gamma^{\mu}\partial_{\mu}\psi-\frac{g}{2}(\bar{\psi}\psi)^2+\frac{g}{2}\left(\frac{\sigma}{g}-\bar{\psi}\psi\right)^2
\nonumber\\
&+&\frac{1}{2}(h V_{\mu}-\partial_{\mu}\sigma)(h V^{\mu}-\partial^{\mu}\sigma)+\bar{\eta}\psi+\bar{\psi}\eta\bigg{)}\bigg{]}
\nonumber\\
&=&\int [d\bar{\psi}][d\psi][ d\sigma][dV_{\mu}]\exp\bigg{[}i\int d^4x \bigg{(}i\bar{\psi}\gamma^{\mu}\partial_{\mu}\psi-\sigma\bar{\psi}\psi +\frac{\sigma^2}{2g}+
\nonumber\\
&+&\frac{1}{2}(h^2 V_{\mu}V^{\mu}-h\partial_{\mu}\sigma V^{\mu}-hV_{\mu}\partial^{\mu}\sigma +\partial_{\mu}\sigma\partial^{\mu}\sigma)+\bar{\eta}\psi+\bar{\psi}\eta\bigg{)}\bigg{]},
\label{M44}
\end{eqnarray}
where $h$ is a constant. On setting $h$ to zero, the integrand becomes independent of $V_{\mu}$, the $V_{\mu}$ path integration becomes irrelevant, and $Z(\bar{\eta}, \eta)$ reduces to 
\begin{eqnarray}
Z(\bar{\eta}, \eta)&=&\int [d\bar{\psi}][d\psi][ d\sigma]\exp\bigg{[}i\int d^4x \bigg{(}i\bar{\psi}\gamma^{\mu}\partial_{\mu}\psi-\sigma\bar{\psi}\psi +\frac{\sigma^2}{2g}+
\frac{1}{2}\partial_{\mu}\sigma\partial^{\mu}\sigma+\bar{\eta}\psi+\bar{\psi}\eta\bigg{)}\bigg{]}.
\nonumber\\
\label{M45}
\end{eqnarray}
As we see, the fermion Green's functions of the NJL theory are given as the fermion Green's functions of a scalar field theory whose action is precisely the NJL mean field action as evaluated in coherent states in which a $\sigma$ kinetic energy term is generated \cite{footnote3}. In consequence, the perturbative expansions in the two theories are in one to one correspondence. However, in this scalar field theory there is no source term $J\sigma$ for the scalar field (in a true fundamental Higgs field Lagrangian there would be such a source term), and thus the scalar field theory only generates Green's functions with external fermion legs and does not generate any Green's functions with external scalar field legs. Thus in the dynamical Higgs case one can generate the fermion Green's functions using an elementary Higgs field theory in which the elementary Higgs field only role is to contribute internally in Feynman diagrams and never to appear in any external legs. In such a case, the all-order iteration of internal $\sigma$ exchange diagrams then generates the dynamical Higgs and Goldstone poles in $T_{\rm S}(q^2,M)$ and $T_{\rm P}(q^2,M)$. Moreover, as we had noted above, the dynamically generated Higgs boson will be an above-threshold resonance and have a nonzero width. Since a fundamental Higgs boson can contribute both off shell and on shell while the dummy field $\sigma$ can only contribute off shell, it is in the on-shell behavior in the Higgs sector and through the width that it would acquire that one may ultimately be able to distinguish between an elementary Higgs boson and a dynamical one.

\section{The Moral of the Story}

With the vacuum of quantum field theory being a dynamical one,  in a sense Einstein's ether has reemerged. Only it has reemerged not as the mechanical ether of classical physics that was excluded by the Michelson-Morley experiment, but as a dynamical, quantum-field-theoretic one full of Dirac's negative energy particles, an infinite number of such particles whose dynamics can spontaneously break symmetries. The type of physics that would be taking place in this vacuum depends on how symmetries are broken, i.e. on whether the breaking is by  elementary Higgs fields or by dynamical ones. If the symmetry is broken by an elementary Higgs field, then the Higgs boson gives mass to fundamental gauge bosons and fermions alike. However, if the breaking is done dynamically, then it is the structure of an ordered vacuum itself that generates masses, with the mass generation mechanism in turn then producing the Higgs boson. In the dynamical case then mass produces Higgs rather than Higgs produces mass. In the dynamical case we should not be thinking of the Higgs boson as being the ``god particle". Rather, if anything, we should be thinking of the  vacuum as being the ``god vacuum".


\begin{thebibliography}{99}
\bibitem{footnote3} The author is indebted to Dr. Joshua Erlich for asking a helpful question regarding the $\sigma$  field kinetic energy term.
\end{thebibliography}
\end{document}